\documentclass[a4paper,11pt]{article}
\usepackage{pos}
\usepackage{babel}[eng]
\usepackage{inputenc}[UTF-8]
\usepackage{amsmath}
\usepackage{physics}
\usepackage{xcolor}
\usepackage{bbold}
\usepackage{tikz}
\usetikzlibrary{positioning,decorations.markings,calc}
\usepackage{graphicx}
\usepackage{hyperref}
\usepackage{subfig}
\graphicspath{{.}}
\DeclareMathOperator{\de}{\partial}
\renewcommand{\vec}[1]{\mathbf{#1}}
\renewcommand{\bar}{\overline}

\title{Ground-State Extraction of Heavy-Light Meson Semileptonic Decay Form Factors}

\author*[a,b]{A.~D'Anna}
\author[c,d]{A.~Conigli}
\author[e]{P.~Fritzsch}
\author[f]{A.~Gérardin}
\author[g]{J.~Heitger}
\author[a,b]{G.~Herdoíza}
\author[a,b,h]{N.~Husung}
\author[h]{S.~Kuberski}
\author[a,b]{C.~Pena}
\author[i]{H.~Simma}

\affiliation[a]{Instituto de Física Teórica UAM-CSIC, Universidad Autónoma de Madrid, Cantoblanco, Madrid, Spain}
\affiliation[b]{Dpto. de Física Teórica, Universidad Autónoma de Madrid, Cantoblanco, Madrid, Spain}
\affiliation[c]{Helmoltz Institute Mains, Johannes Gutenberg University, Mainz, Germany}
\affiliation[d]{GSI Helmholtz Center for Heavy Ion Research, Darnstadt, Germany}
\affiliation[e]{School of Mathematics, Trinity College, College Green, Dublin, Ireland}
\affiliation[f]{Aix-Marseille Université, Université de Toulon, CNRS, CPT, Marseille, France}
\affiliation[g]{Institut f\"ur Theoretische Physik, Universit\"at M\"unster, M\"unster, Germany}
\affiliation[h]{Theoretical Physics Department, CERN, 1211 Geneva 23, Switzerland}
\affiliation[i]{John von Neumann-Institut für Computing NIC, Deutsches Elektronen-Synchrotron DESY, Zeuthen, Germany}

\emailAdd{antonino.danna@estudiante.uam.es}

\abstract{We discuss the extraction of heavy-light pseudo-scalar to light pseudo-scalar decay form factors from finite time correlation functions. We place particular emphasis on the contamination from excited states employing summed ratios and input from chiral perturbation theory. The analysis is  performed on four CLS ensembles with $N_f = 2+1$ flavours of $\mbox{O}(a)$-improved Wilson fermions (presently) at the  $\mathrm{SU}(3)$-symmetric point with relativistic heavy-quark masses in the  charm region and above. The study presented here is part of the analysis aimed at the computation of the  $B \to \pi \ell \nu$ and $B_s \to K \ell \nu$ semileptonic form factors, combining the continuum-limit relativistic results with static-limit calculations.}
\FullConference{The 42nd International Symposium on Lattice Field Theory (LATTICE2025)\\
  2-8 November 2025\\
  Tata Institute of Fundamental Research, Mumbai, India\\}

\begin{document}
\maketitle
\section{Introduction}
\label{Sec:introduction}
High-precision measurements of the CKM matrix elements are crucial in constraining the Standard Model. In this context, computing the semileptonic decay form factors associated with the $B \to \pi \ell \nu$  and $B_s\to K \ell \nu $ decays from first principles enables the determination  of the CKM matrix element $|V_{ub}|$.
Here we present a study of the extraction of the relevant ground-state matrix elements based on the summed ratio method \cite{Maiani-etal,Gusken-etal,B-D-S-summed-ratio,Bahr:2019eom}. Moreover, we discuss the treatment of excited-state contamination using input from Heavy Meson Chiral Perturbation Theory (HMChPT)~\cite{Broll-thesis,Bar:2023sef}.

\begin{table}
  \centering
  \begin{tabular}{|r|cccccclc|}
    \hline
    id   &  $\beta$ & $a$ [fm] & $L$ [fm] & $T$ [fm] & $x_0^{\rm src}$ [fm] & $m_\pi L$ & $N_{\rm conf}$      &  $N_{h}$\\  \hline
    H101 &  $3.40$  & $0.085$  & $2.7$    & $8.2$    & $2.6$              & $5.8$     & $1007$, $1009$      & $6$     \\
    H400 &  $3.46$  & $0.076$  & $2.4$    & $7.3$    & $2.1$              & $5.2$     & $505$, $540$        & $6$     \\
    N202 &  $3.55$  & $0.063$  & $3.0$    & $8.1$    & $2.5$              & $6.5$     & $899$, $1003$       & $5$     \\
    J500 &  $3.85$  & $0.039$  & $2.5$    & $7.4$    & $2.3$              & $5.2$     & $789$, $655$, $431$ & $5$     \\  \hline
  \end{tabular}
  \caption{Lattice parameters. For each ensemble at the $\mathrm{SU}(3)$ symmetric point (${m_\pi = m_K \approx 410 \text{ MeV}}$), we simulate $N_{h}$ heavy quarks with masses $m_c^{\rm RGI} \lesssim m_h^{\rm RGI} \lesssim 2m_c^{\rm RGI}$, and pions with momentum ${|\vec p_\pi| \approx 0,\, 480,\, 765 \text{ MeV}}$. The two- and three-point functions share the same source position, quoted in the column $x_0^{\rm src}$ as distance from the closest boundary, and are computed using ten stochastic sources. In the three-point function computations, five distinct source-sink separations in the range $1.4 \text{ fm} \lesssim t_s \lesssim 2.8 \text{ fm}$ are considered. In the column $N_{\rm conf}$ we quote the number of configurations for each replica for a given ensemble.}
  \label{Tab:lattices}
\end{table}

The study presented here is part of our ongoing effort to compute the semileptonic decay form factors, $f_{+,0}(q^2)$ defined in eq.~\eqref{Eq:f+0} below, associated to the $B\to \pi \ell\nu$ and $B_s \to K \ell\nu$ semileptonic decays at the physical point. To this purpose, we implement the strategy presented in Ref.~\cite{strategy-for-B-physics} and already employed to extract the $b-$quark mass in Refs.~\cite{AC-proceeding, AC-europlex}, and more recently in Ref.~\cite{Simon-Talk}. This approach relies on the interpolation to the $b-$sector of  the relativistic theory  combined with the static limit of  Heavy Quark Effective Theory (HQET)~\cite{Sommer:2015hea}. Suitable observables, usually ratios or logs of the target observables, are constructed in such a way that neither matching nor renormalization is required. As a result these observables are free from logarithmic corrections at leading order.
In our case, where we  use the HQET parametrization of the matrix element relevant to the semileptonic decays, we aim to extract the form factors $h_\perp$ and $h_\|$, defined in eq.~\eqref{Eq:M_HQET} below, at $m_h = m_b$. In the case of $h_\perp$ (a similar treatment applies to $h_\|$), we employ two auxiliary observables $\phi$ and $\tau_{\perp}$ defined as

\begin{equation}
  \label{Eq:tau_perp}
  \phi = \ln(L_{\rm ref}^{3/2}\hat f_{B^*}), \quad
  \tau_{\perp}(E_P)= \ln(\frac{E_P h_\perp(E_P)}{L_{\rm ref} \hat f_{B^*}}),
\end{equation}
where $\hat f_{B^*}$ is the $B^*-$meson decay constant in the non-relativistic normalization and $E_P$ is the energy of the pseudoscalar meson in the final state. After computing $\tau_\perp$ in the static point of HQET and in the relativistic theory at heavy quark masses in the charm region, we interpolate it to the physical $b-$quark mass. Regarding $\phi$ in eq.~\eqref{Eq:tau_perp}, an additional step-scaling procedure in volume is required to connect $\phi(L_1)$, estimated at small volume ($L_1 = 0.5 \text{ fm}$) where the relativistic $b-$quarks can be simulated, to the large-volume ensembles generated by the Coordinated Lattice Simulation (CLS) \cite{CLS} initiative as
\begin{equation}
  \phi = \phi(L_1) + [\phi(L_2) - \phi(L_1)] + [\phi_{\rm CLS} - \phi(L_2)],
\end{equation}
where $L_2 = 2L_1$. The step-scaling functions, $[\phi(L_2)-\phi(L_1)]$ and $[\phi_{\rm CLS} - \phi(L_2)]$, extrapolated to the continuum, are independently interpolated to the physical $b-$quark mass before combining them with $\phi(L_1)$ to obtain $\phi$. For the step-scaling functions, as well as $\tau_\perp$, no renormalization factors nor matching coefficients between the static point of HQET and the relativistic theory are required, and the observables are free from log terms at order $\mbox{O}(1/m_b)$.

Once $\phi$ and $\tau_\perp$ have been determined at $m_h = m_b$, the form factor $h_\perp(E_P)$ can be extracted through the following expression
\begin{equation}
  \ln(E_P L_{\rm ref}^{1/2}h_{\perp}(E_P)|_{m_h=m_b}) = \phi|_{m_h=m_b} + \tau_\perp(E_P)|_{m_h=m_b}.
\end{equation}

\begin{sloppypar}
  In the study presented here, we concentrate on the extraction of the form factor $h_\perp(E_P)$ needed to construct $\tau_\perp$ in eq.~\eqref{Eq:tau_perp}.  We use four large-volume CLS ensembles (see Tab.~\ref{Tab:lattices})  which are located at the $\mathrm{SU}(3)$ symmetric point (${m_\pi = m_K \approx 410 \text{ MeV}}$), with $N_f=2+1$ $\mbox{O}(a)$-improved Wilson Fermions and open boundary condition in time.
On these ensembles we measure two- and three-point functions of heavy-light mesons with heavy relativistic quark masses in the range $m_c^{\rm RGI} \lesssim m_h^{\rm RGI} \lesssim 2 m_c^{\rm RGI}$.
\end{sloppypar}

\section{$B\to \pi \ell \nu$ Semileptonic Decay Form Factors}
The differential decay rate of the $B \to \pi \ell \nu$ semileptonic decay can be written as:
\begin{equation}
  \begin{split}
    \frac{\dd \Gamma}{\dd q^2} =& \frac{G_F^2 |\eta_{\rm EW}|^2 |V_{ub}|^2 }{ 24 \pi^3 } \frac{(q^2-m_\ell^2)^2\sqrt{E_\pi^2-m_\pi^2}}{q^4 m_B^2}\\
                                &\times\Bigg[\left(1 + \frac{ m_\ell^2 }{ 2 q^2 } \right) m_{B}^2 (E_\pi^2 -m_\pi^2) |f_+(q^2)|^2 + \frac{ 3m_\ell^2 }{ 8 q^2 } (m_{B}^2-m_\pi^2) |f_0(q^2)|^2 \Bigg],
  \end{split}
\end{equation}
where $q = p_{B} - p_\pi$, $\eta_{EW}$ is the short-distance electroweak correction factor, $m_\ell$ is the lepton mass, and the form factor $f_+(q^2)$ and $f_0(q^2)$ are related to the matrix element describing the process
\begin{equation}
  \label{Eq:f+0}
  \langle \pi(\vec p_\pi) |\hat V_\mu|B (\vec p_B)\rangle = f_+(q^2)\Big(p_{B,\mu} + p_{\pi,\mu} - \frac{m_{B}^2 -m_\pi^2 }{q^2}q_\mu\Big) + f_0( q^2) \frac{m_{B}^2 - m_\pi^2}{q^2}q_\mu,
\end{equation}
where $\hat V_\mu = Z_V V_\mu$ is the renormalized local heavy-light vector current, and $Z_V$ is the corresponding renormalization factor \cite{ZV,ZV2}.
On the lattice, it is convenient to use the HQET parametrization, in which the form factor, $h_\perp(E_\pi)$ and $h_\|(E_\pi)$, are defined as
\begin{equation}
  \label{Eq:M_HQET}
  \langle \pi |\hat V_\mu|B \rangle = \sqrt{2m_{B}}\left( v_\mu h_\|(E_\pi) + p_{\mu}^\perp h_\perp(E_\pi)\right),
\end{equation}
where $v_\mu$ is the $B-$meson $4$-velocity and $p^\perp_\mu = p_{\pi,\mu} - (p_\pi\cdot  v)v_\mu$. In the $B-$meson rest frame, the form factors $f_{+,0}(q^2)$ are recovered by
\begin{align}
  f_+(q^2) =& \frac{1}{\sqrt{2m_{B}}} \left[h_\|(E_\pi) + (m_{B} - E_\pi) \, h_\perp(E_\pi) \right], \\
  f_0(q^2) =& \frac{\sqrt{2m_{B}}}{m_{B}^2 -  m_\pi^2}\left[(m_{B} - E_\pi)\, h_\|(E_\pi) + (E_\pi^2 - m_\pi^2)\, h_\perp(E_\pi)\right].
\end{align}
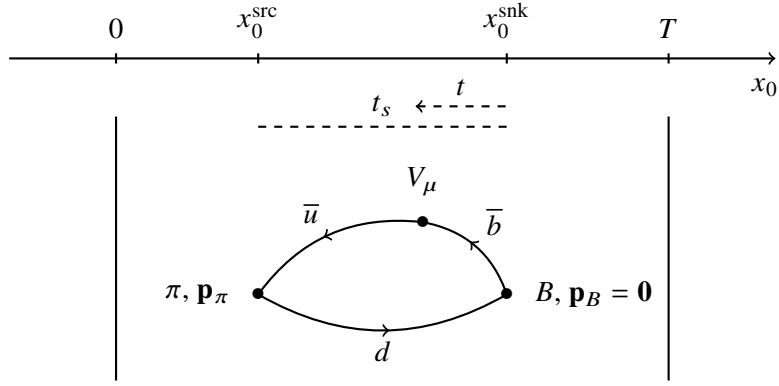
\begin{figure}[t]
  \centering
  \begin{tikzpicture}
    \node[]                   (e1)                       {};
    \node[label={90:$0$}]     (a1) [right=1   cm of e1]  {};
    \node[label={90:$x_0^{\rm src}$}]   (b1) [right=1.6 cm of a1]  {};
    \node[label={90:$x_0^{\rm snk}$}]   (c1) [right=3   cm of b1]  {};
    \node[label={90:$T$}]     (d1) [right=7   cm of a1]  {};
    \node[label={270:$x_0$}]  (e2) [right=1   cm of d1]  {};
    \node[]                   (a2) [below=0.5 cm of a1]  {};
    \node[]                   (b2) [below=0.5 cm of b1]  {};
    \node[]                   (h3) [right=1.8 cm of b2]  {};
    \node[]                   (c2) [right=3   cm of b2]  {};
    \node[]                   (d2) [below=0.5 cm of d1]  {};
    \node[label={180:$\pi$, $\vec{p}_\pi$}]  (b3) [below=2.0 cm of b2]  {$\bullet$};
    \node[label={0:$B$, $\vec p_B = \vec 0$}] (c3) [below=2.0 cm of c2]  {$\bullet$};
    \node[]                   (a3) [below=4   cm of a1]  {};
    \node[]                   (d3) [right=7   cm of a3]  {};
    \node[]                   (h1) [above=0.6 cm of b3]  {};
    \node[label={90:$V_\mu$}] (h2) [right=1.8 cm of h1]  {$\bullet$};
    \begin{scope}[thick,decoration={
        markings,
        mark=at position 0.5 with {\arrow{<}}}
      ]
      \draw[bend left,postaction={decorate}]  (b3.center) to node [auto]{$\bar u$} (h2.center);
      \draw[bend left,postaction={decorate}]  (h2.center) to node [auto]{$\bar b$} (c3.center);
      \draw[bend left,postaction={decorate}]  (c3.center)   to node [auto]{$d$} (b3.center);
      \draw[->] (e1.west) to  (e2.east);
      \draw[-] (a2.center) to (a3.center);
      \draw[-] (d2.center) to (d3.center);
      \draw[dashed] (b2.south) to node[above]{$t_s$} (c2.south);
      \draw[<-,dashed] (h3.north) to node[above]{$t$} (c2.north);
      \draw[-] ($(a1.north)!0.25!(a1.south)$) to ($(a1.north)!0.75!(a1.south)$);
      \draw[-] ($(b1.north)!0.25!(b1.south)$) to ($(b1.north)!0.75!(b1.south)$);
      \draw[-] ($(c1.north)!0.25!(c1.south)$) to ($(c1.north)!0.75!(c1.south)$);
      \draw[-] ($(d1.north)!0.25!(d1.south)$) to ($(d1.north)!0.75!(d1.south)$);
    \end{scope}
  \end{tikzpicture}
  \caption{Schematic representation of a three-point function on the lattice. The ensembles under consideration have open boundary condition in time and the boundaries are located at $x_0=0$ and $x_0 = T$. The source position, $x_0^{\rm src}$, is kept fixed in the two- and three-point functions of a given ensemble. For the three-point functions, five distinct values of the sink position $x_0^{\rm snk}$ are considered. The injection time of the vector current varies starting from the $B$-meson interpolating operator, at $t=0$, to the $\pi$ interpolating operator at $t = t_s = x_0^{\rm snk}- x_0^{\rm src}$.}
  \label{Fig:3pt-diagram}
\end{figure}
The semileptonic decay under study is described by the three-point function (Fig.~\ref{Fig:3pt-diagram})
\begin{equation}
  \label{Eq:3pt}
    C_\mu(t,t_s, \vec p_\pi) = \langle \mathcal P(t_s,\vec p_\pi) V_\mu(t) \mathcal{B}(0,\vec 0)\rangle ,
\end{equation}
where $\mathcal P$ and $\mathcal{B}$ are the interpolating operators with the quantum numbers of a pion  and $B-$meson, respectively, $t$ is the injection time of the vector current $V_\mu$, and $t_s = x_0^{\rm snk}-x_0^{\rm src}$ is the source-sink time separation. At asymptotic time separations, the matrix element $\langle \pi |V_\mu|B\rangle$ can be extracted using the following ratios of three- and two-point functions \cite{exclusive_Bs_to_k,Hadronic_ff,Bahr:2019eom}
\begin{align}
  \label{Eq:RI}
  \mathcal R_\mu^{I} (t,t_s,\vec p_\pi)   =& \sqrt{4E_\pi m_{B}} \frac{C_\mu(t,t_s, \vec p_\pi)}{\sqrt{C_{B}(t,\vec 0)C_\pi(t_s - t,\vec p_\pi)}} \exp(\frac{1}{2} E_\pi (t_s - t) +\frac{1}{2} m_{B} t),\\
  \label{Eq:RII}  \mathcal R_\mu^{II} (t,t_s,\vec p_\pi)   =& \sqrt{4E_\pi m_{B} \frac{C_\mu(t,t_s, \vec p_\pi)C_\mu(t_s-t,t_s, \vec p_\pi)}{C_B(t_s,\vec 0)C_\pi(t_s,\vec p_\pi)}},\\
  \label{Eq:RIII}  \mathcal R_\mu^{III}(t,t_s,\vec p_\pi)   =& \sqrt{4E_\pi m_{B}}\frac{C_\mu(t,t_s, \vec p_\pi)}{\sqrt{C_B(t_s,\vec 0)C_\pi(t_s,\vec p_\pi)}}\exp(\frac{m_{B} - E_\pi}{2}(2t-t_s)),
\end{align}
where $E_\pi$ is estimated through the lattice dispersion relation using the pion mass extracted from the pion two-point function  $C_\pi(t,\vec 0)$. Likewise, $m_B$ is extracted from the $B-$meson two-point function $C_B(m_B,\vec 0)$.
In the limit $0\ll t \ll t_s$, we can relate these ratios to the form factors as ($X=I,II,III$)
\begin{align}
 \label{Eq:h_perp}
  p_{\pi,k} h_\perp(E_\pi) =&  \frac{Z_V}{\sqrt{2m_{B}}} \lim_{0\ll t \ll t_s} \mathcal R_k^X(t,t_s,\vec p_\pi),\\
  h_\|(E_\pi) =& \frac{Z_V}{\sqrt{2 m_{B}}} \lim_{0\ll t \ll t_s} \mathcal R_0^X(t,t_s,\vec p_\pi).
\end{align}
It is worth noting that $h_\perp(E_\pi)$ is only accessible at non-zero momentum and that eq.~\eqref{Eq:h_perp} is valid independently for each spatial direction $k$.
Given that we simulate the pion with equal momentum in the three spatial directions, we replace the index $k$ by $\perp$ to indicate an average over those directions. For example, to extract $h_\perp(E_\pi)$ we use the ratio $\mathcal{R}^X_{\perp} = \frac{1}{3}\sum_{k=1}^3 \mathcal{R}_k^X$, for $X = I,II,III$.

As we increase the source-sink separation $t_s$, an exponential degradation of the signal is observed, thus limiting us to small value of $t_s$. Under these conditions, the contamination from excited states are not sufficiently suppressed and they constitute an important part of the signal. In particular, it has been shown that the leading excited-state contribution affecting the ratio $\mathcal{R}_\mu^{III}$ is of $\mbox{O}\left(e^{-t_s \Delta/2}\right)$, where  $\Delta = \min\{\Delta_\pi,\Delta_B\}$  is the smallest energy gap between the ground state and the first excited state \cite{Bahr:2019eom}.

The top panels of Fig.~\ref{Fig:summed_ratios_and_hperp} show the effective form factor $h_\perp(t,t_s, E_\pi)$ computed using the different ratios in eqs.~(\ref{Eq:RI}-\ref{Eq:RIII}). The ratios visibly change as we increase the source-sink separations. In particular, at $t_s\approx 1.6\text{ fm}$, the excited states are still heavily present, and the ratios cannot saturate into a plateau.
As we increase the source-sink separation, the ratios visibly change due to the reduced presence of the excited states.
However, even at $t_s\approx 2.7 \text{ fm}$ we do not observe unambiguous evidences of a stability region corresponding to a ground state.

\section{Summed Ratios}
\label{Sec:summed_ratios}
The summed ratios \cite{Maiani-etal,Gusken-etal,B-D-S-summed-ratio,Bahr:2019eom} can be employed to mitigate the excited-state contaminations in the extraction of the matrix elements. A summed ratio is defined as:
\begin{equation}
  \label{Eq:Summed ratio}
  S_\mu^X(t_s,\vec p_\pi) =  \sum_{t=0}^{t_s} \mathcal R_\mu^X(t, t_s,\vec p_\pi) = K + \mathcal{M}_\mu(\vec p_\pi) t_s + \text{ excited states, }
\end{equation}
where $K$ is a constant and $\mathcal{M_\mu}$ the ground-state matrix element that the ratio isolates. In our case $\mathcal{M}_\mu = \langle \pi|V_\mu|B\rangle$ is defined in eq.~\eqref{Eq:M_HQET}. From the $S_\mu^X$, we can extract $\mathcal M_\mu$ by taking a derivative with respect to $t_s$
\begin{equation}
  \label{Eq:M_mu_SR}
  \mathcal{M}_\mu^X(\vec p_\pi) = \de_{t_s}S_\mu^X(t_s,\vec p_\pi) = \mathcal{M}_\mu(\vec p_\pi) + \text{ excited states. }
\end{equation}
The precise form of the excited‑state contributions is specific to the ratio employed. For example, the leading excited-state contributions to $\mathcal{M}_\mu^{III}$ is of $\mbox{O}(t_s \Delta e^{-\Delta t_s})$ \cite{Bahr:2019eom}. The excited-state contributions are mitigated by the summed ratio as compared to the contributions from $\mathcal{R}^{III}_\mu$, which are  of $\mbox{O}(e^{-t_s\Delta/2})$.

\begin{figure}
  \centering
  \includegraphics[width=\textwidth]{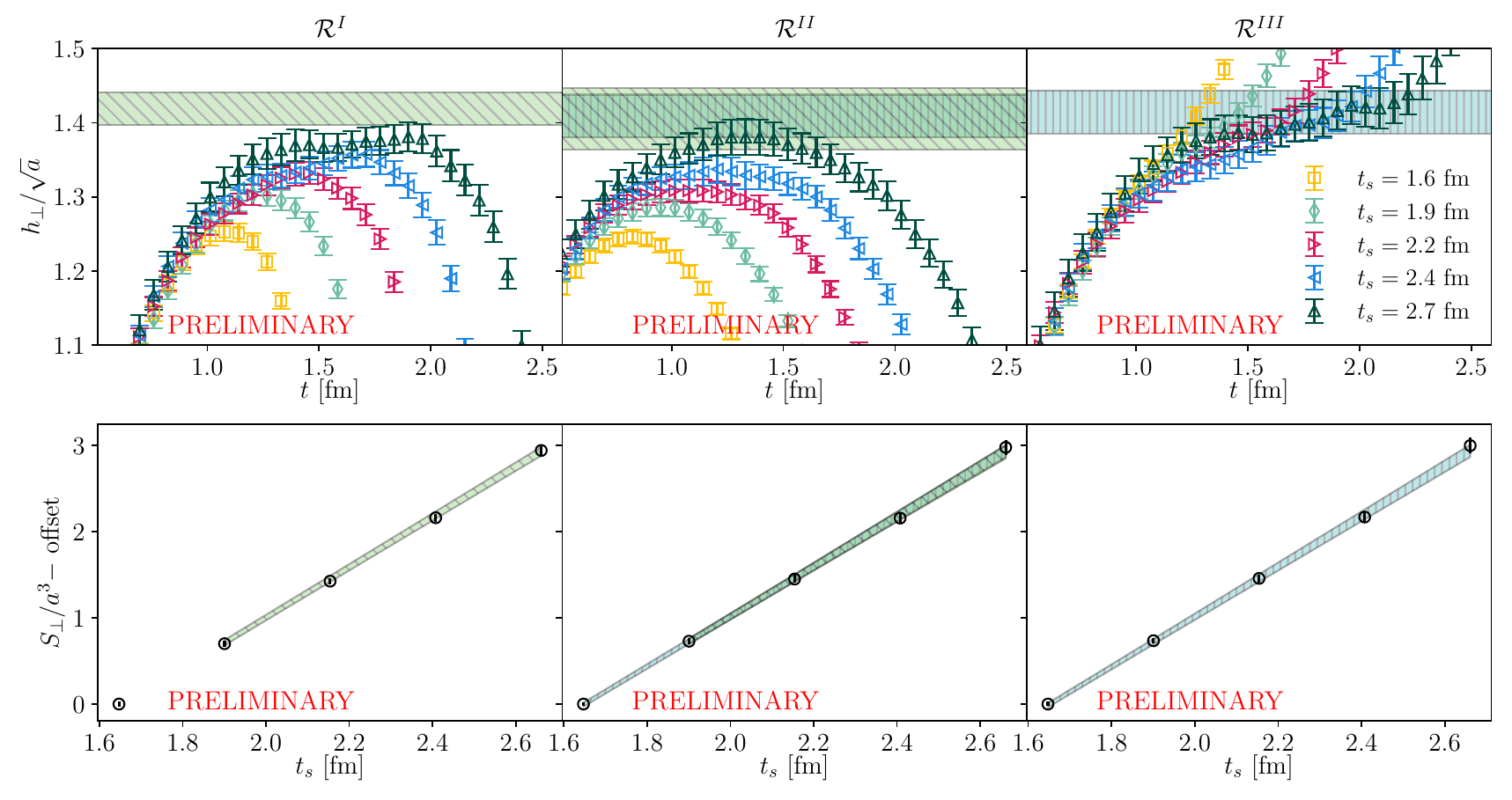}
  \caption{Top row: Effective form factor  $h_\perp(t,t_s,E_\pi)$ corresponding to eq.~\eqref{Eq:h_perp} computed using the ratios in eqs.~(\ref{Eq:RI}-\ref{Eq:RIII}). The $B-$meson interpolating operator sits at $t=0$. Bottom row: Fit to the summed ratio $S_\perp(t_s,\vec p_\pi)$ in eq.~\eqref{Eq:Summed ratio}. The points are shifted in such a way that $S_\perp$ at $t_s \approx 1.6\text{ fm}$ is at $0$. The coloured bands are the result of a correlated linear fit to $S_\perp$ to extract the ground-state matrix element. The bands in the bottom row are drawn over the points that enter the fit. The horizontal bands in the top row correspond to the form factors obtained from the fit shown below. Bands of the same colour share the same $t_s$ fit range.  Only fits with $\text{p-value}\ge 0.05$ are shown. The data correspond to the ensemble N202 ($a = 0.063 \text{ fm}$) with heavy quark mass $m_h^{\rm RGI} = 0.93 m_c^{\rm RGI}$, at $|\vec{p}_\pi| = 479 \text{ MeV}$.}
  \label{Fig:summed_ratios_and_hperp}
\end{figure}

To extract the matrix element, we fit the summed ratios with a linear function following the corresponding terms in eq.~\eqref{Eq:Summed ratio}. Since smaller source-sink separations are expected to suffer the largest excited‑state contributions, we explore systematic effects by imposing the cut $t_s >2.0 \text{ fm}$. In Fig.~\ref{Fig:summed_ratios_and_hperp}, we show the fits with $\text{p-value}\ge 0.05$.
The form factor values extracted from the various ratios agree within uncertainties. Using $\mathcal{R}_\mu^I$ gives a $1-2$\% statistical precision on $h_\perp$, while $\mathcal{R}_\mu^{II}$ and $\mathcal{R}_\mu^{III}$ yield $2-3$\%. A detailed assessment of systematic uncertainties is in progress.

\section{$B^*\pi$ Excited-State Contributions}
Both a HMChPT analysis~\cite{Bar:2023sef, Broll-thesis} and a variational method~\cite{Barca-CurrentEnhanced} predict that the excited-state contributions to the three-point function related to $h_\perp$, $C_k$ in eq.~\eqref{Eq:3pt}, are dominated at intermediate source-sink separations by one volume-enhanced $B^*\pi$ state. Following the argument of Ref.~\cite{Barca-CurrentEnhanced}, such volume-enhanced contributions arise from the quark-line disconnected diagrams in which the $B^*-$meson couples with the spatial vector current while the pion propagates from source to sink. On the other hand, the diagram responsible for the volume enhanced contribution vanishes in $h_\|$. The $B^*\pi$ contributions to $h_\perp$ have been computed in HMChPT at NLO in the static approximation~\cite{Bar:2023sef, Broll-thesis}
\begin{align}
  \label{Eq:Bs_pi_to_Cmu}
  C_k(t,t_s,\vec p_\pi) =& C_k^{\rm gs}(t,t_s,\vec p_\pi)\left(1 - \frac{1+\beta_1 E_\pi/\hat g}{1-\beta_1 E_\pi/\hat g}\exp(-E_\pi t) + \dots\right),\\
  \notag=& C_k^{\rm gs}(t,t_s,\vec p_\pi)\left(1 + \delta C_k(t,t_s,E_\pi)\right),
\end{align}
\begin{sloppypar}
  \noindent
where $\hat g$ is a Leading Order (LO) Low Energy Constant (LEC), $\beta_1$ is a Next-to-Leading Order (NLO) LEC, and $C_k^{\rm gs}$ is the ground-state contribution to the three-point function.
  In the static limit $\hat g = 0.49(3)$ \cite{Bernardoni:2014kla}, and ${\beta_1 = 0.20(4) \text{ GeV}^{-1}}$ \cite{Gérardin_LAT24}. Since at NLO $\delta C_k$ is of order $\mbox{O}(p_\pi)$, while $B^*\pi$ excited-state contribution coming from $C_B$ and $m_B$ are of order $\mbox{O}(p^2_\pi)$ \cite{Broll-thesis}, the dominant contributions to $h_\perp$ are those coming from $\delta C_k$. It is worth noting that the HMChPT description, being a low energy effective theory, requires that the low energy pions dominate the contribution in $\delta C_k$. Consequently, the HMChPT expansion is expected to be reasonably well-behaved for sufficiently large time separations, typically~$t \gtrsim 1.3 \text{ fm}$~\cite{Broll-thesis,Bar:2023sef}.
\end{sloppypar}
\begin{figure}[t]
  \centering
  \includegraphics[width=\textwidth]{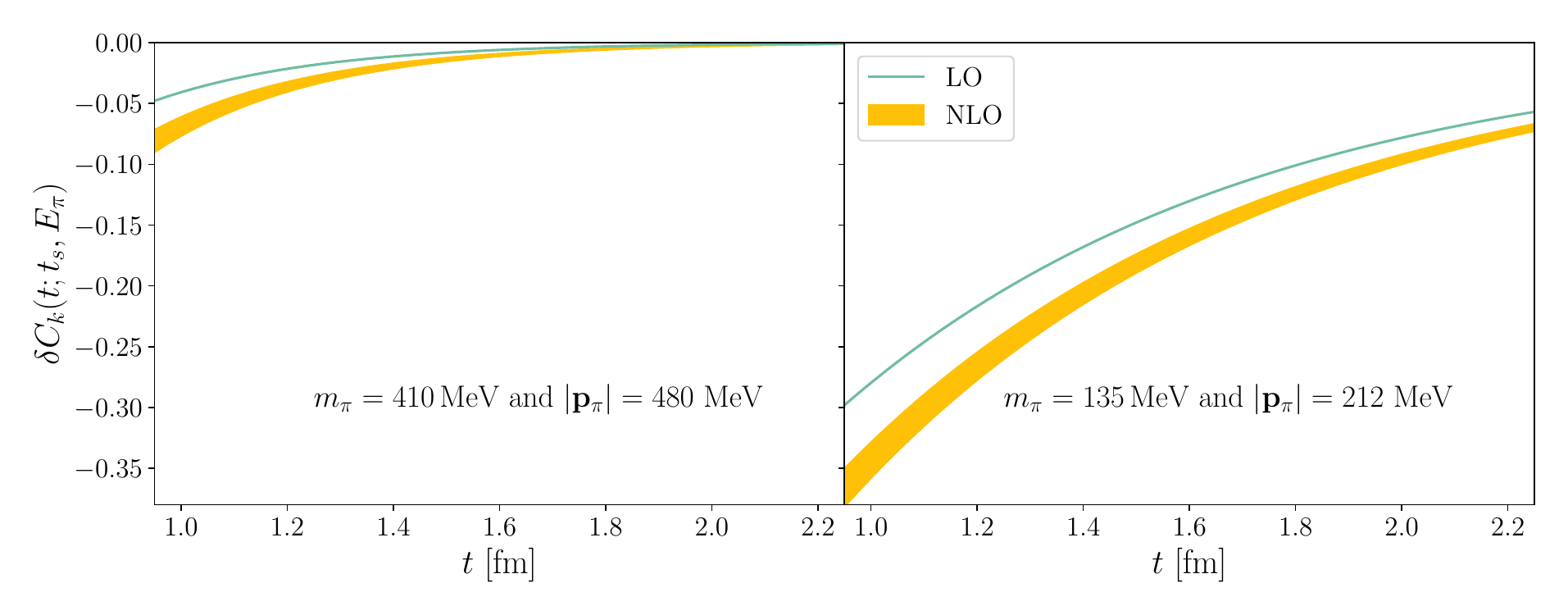}
  \caption{$B^*\pi$ contributions to $C_k$ based on HMChPT \cite{Broll-thesis,Bar:2023sef}, see eq.~\eqref{Eq:Bs_pi_to_Cmu}. The left panel shows $\delta C_k$ for the $\mathrm{SU}(3)$ symmetric point, the right panel correspond to  the physical pion mass. In both cases, $m_\pi L = 4$. The error band comes from $\beta_1 = 0.20(4) \text{ GeV}^{-1}$ \cite{Gérardin_LAT24} and $\hat g=0.49(3)$ \cite{Bernardoni:2014kla}. Comparing the two plots, we clearly see how the contributions coming from the $B^*\pi$ states increase as we approach the physical pion mass.}
  \label{Fig:Dhperp}
\end{figure}

For symmetric point ensembles ($m_\pi = m_K \approx 410 \text{ MeV}$) and the smallest non-zero pion energy under consideration ($|\vec p_\pi| = 479 \text{ MeV}$) the contributions coming from the $B^*\pi$ excited states based on HMChPT are expected to stay below $10$\% of the total correlator for $t \gtrsim 1 \text{ fm}$ and well below $5$\% at $t \ge 1.3 \text{ fm}$. As we get closer to the physical pion mass and/or the pion energy is reduced, the contributions coming from $B^*\pi$ become more important. For example, in the right panel of Fig.~\ref{Fig:Dhperp}, we illustrate the case with physical pions at $|\vec p_\pi| = 212 \text{ MeV}$, that is the smallest non-zero momentum compatible with the momentum quantisation rule when assuming $m_\pi L = 4$. We observe that the $B^*\pi$ states account for $\sim 23$\% of the total three-point correlator at $t \approx 1.3 \text{ fm}$, and  still contribute at $\sim 10$\% at $t\approx 2.0 \text{ fm}$, where we observe excited-state contributions coming from the pion operator at the largest source-sink separation considered, i.e. $t_s\approx 2.7\text{ fm}$.
\begin{figure}[t]
  \centering
  \includegraphics[width=0.75\textwidth]{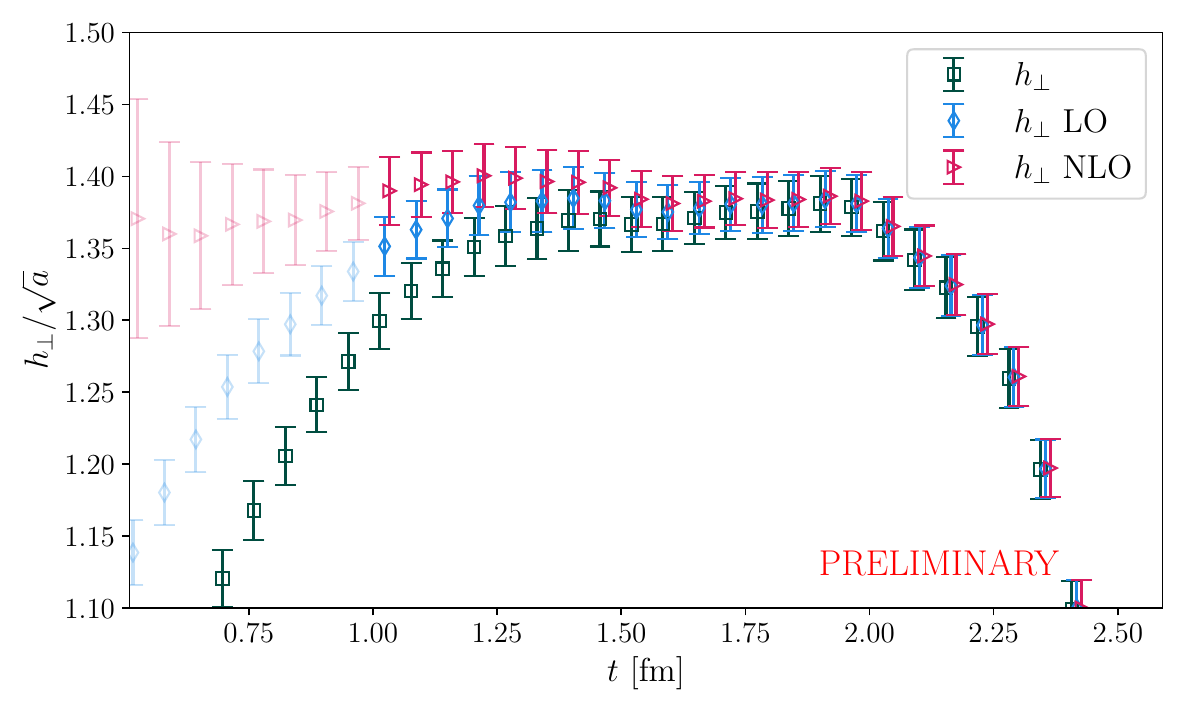}
  \caption{Effective form factor $h_\perp(t,t_s,E_\pi)$ corresponding to  eq.~\eqref{Eq:h_perp} computed using $\mathcal{R}_k^I$ in eq.~\eqref{Eq:RI}, with $B^*\pi$ state subtraction using HMChPT as in eq.~\eqref{Eq:Bs_pi_to_Cmu}. At LO the LEC $\beta_1$ vanishes, while at NLO  $\beta_1 = 0.20(4) \text{ GeV}^{-1}$ \cite{Gérardin_LAT24} is employed. The data points are slightly displaced horizontally for better visibility. Points with higher opacity in the region $t\gtrsim 1.0\text{ fm}$ identify the region where HMChPT is expected to be better behaved. The data correspond to the ensemble N202 ($a = 0.063 \text{ fm}$) with heavy quark mass $m_h^{\rm RGI} = 0.93 m_c^{\rm RGI}$, at $|\vec{p}_\pi| = 479 \text{ MeV}$ and $t_s = 2.7 \text{ fm}$.}
  \label{Fig:Bspi_effects}
\end{figure}
\begin{figure}[t]
  \centering
  \includegraphics[width=\textwidth]{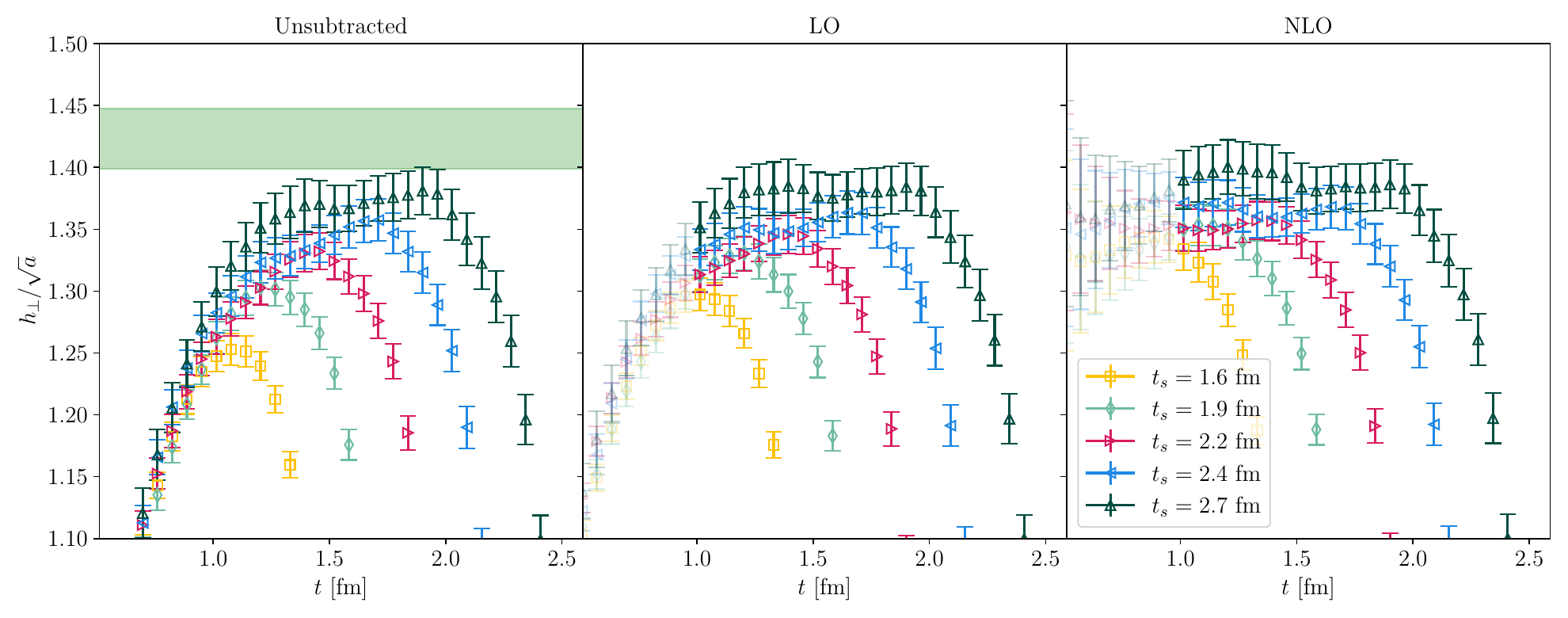}
  \caption{Effective form factor $h_\perp(t,t_s,E_\pi)$ corresponding to eq.~\eqref{Eq:h_perp} computed using $\mathcal{R}^I_k$ in eq.~\eqref{Eq:RI}. Left:~unsubtracted $h_\perp$. The green band is the result of the summed ratio method shown in Fig~\ref{Fig:summed_ratios_and_hperp}. Middle: $h_\perp$ with the $B^*\pi$ excited states subtracted based on HMChPT at LO, i.e eq.~\eqref{Eq:Bs_pi_to_Cmu} with vanishing $\beta_1$. Right: $h_\perp$ with $B^*\pi$ excited states subtracted at NLO, i.e with $\beta_1 = 0.20(4)$ GeV$^{-1}$ \cite{Gérardin_LAT24}.  Points with higher opacity in the region $t\gtrsim 1.0\text{ fm}$ identify the region where HMChPT is expected to be better behaved. The data correspond to the ensemble N202 ($a = 0.063 \text{ fm}$) with heavy quark mass $m_h^{\rm RGI} = 0.93 m_c^{\rm RGI}$, at $|\vec{p}_\pi| = 479 \text{ MeV}$.}
  \label{Fig:Bspi_compare}
\end{figure}
Using eq.~\eqref{Eq:Bs_pi_to_Cmu}, we first subtract $\delta C_\perp$ from the three-point functions and then compute the ratios with the subtracted correlators. In Fig.~\ref{Fig:Bspi_effects} we show the effect of the  subtraction on the largest source-sink separation on the ensemble N202, at $a = 0.063 \text{ fm}$, with the heavy quark in the charm region and $|\vec p_\pi| = 479 \text{ MeV}$.
We observe that subtracting the $B^*\pi$ contributions mitigates the excited‑state contamination induced by the $B-$meson interpolating operator.
Moreover, as shown in Fig.~\ref{Fig:Bspi_compare}, for smaller source-sink separations, $h_\perp$ increases visibly, getting closer to the data point at  $t_s\gtrsim 2.0 \text{ fm}$.

\section{Conclusion and Outlook}
We have described an analysis of the extraction of the ground-state matrix element relevant for the semileptonic decay form factors associated to the $B\to \pi \ell \nu $ and $B_s\to K \ell \nu$ decays, which in turn give access to the CKM matrix element $|V_{ub}|$.

We employed the summed ratio method \cite{Maiani-etal,Gusken-etal,B-D-S-summed-ratio,Bahr:2019eom}, in which the ratio used to the extract the matrix element is summed over the physical time extent of the correlator. The summed ratios mitigate the excited state contaminations and simplify the extraction of the ground-states matrix element, defined in eq.~\eqref{Eq:M_HQET}, relevant for the determination of the form factors.
Moreover, a linear dependence on the source–sink separation $t_s$ of the three-point correlator is observed as predicted in eq.~\eqref{Eq:Summed ratio}, and the target matrix element is obtained directly from a linear fit.

We have tested this procedure on a set of four ensembles located at the $\mathrm{SU}(3)$ symmetric point ($m_\pi = m_K \approx 410$ MeV) with  heavy quark masses in the charm region and three-point correlators with source-sink separations in the range $1.4 \text{ fm } \lesssim t_s \lesssim 2.8 \text{ fm}$. To extract the ground-state matrix element, we employed the three ratios in eqs.~(\ref{Eq:RI}-\ref{Eq:RIII}) and estimated the form factor $h_\perp(\vec p_\pi)$ using eq.~\eqref{Eq:h_perp}.  The current statistical precision on $h_\perp(E_\pi)$ at $|\vec p_\pi| \approx 480\text{ MeV}$ is at $1-2$\% when using $\mathcal{R}_k^{I}$, and $2-3$\% when using $\mathcal{R}_k^{II}$ and $\mathcal{R}_k^{III}$ (See Fig.~\ref{Fig:summed_ratios_and_hperp}). A dedicated assessment of systematic uncertainties in the summed-ratio approach is in progress.
Furthermore, we discussed the treatment of excited states using input from HMChPT to subtract the $B^*\pi$ excited-state contributions. After the subtraction, we observed that the excited-state contaminations coming from the $B-$meson interpolating operator are milder thus providing additional handle in the ground-state extraction.

We are extending the current study of ground‑state extraction to ensembles closer to the physical pion mass, employing relativistic heavy quarks. In parallel, calculations using static heavy quarks are underway to enable a controlled interpolation to the physical $b-$quark mass. The present analysis is further complemented by a determination of the HMChPT low-energy constants that govern the $B^{*}\pi$ excited states, as well as by a dedicated study incorporating two-hadron interpolators into a variational basis to isolate these states~\cite{Gérardin_LAT24}.

\acknowledgments

We are grateful to our colleagues Oliver B\"ar and Rainer Sommer  for the valuable discussions and comments. This work is partially supported by grants PID2021-127526NB-I00, PID2024-160152NB-I00, and IFT Centro de Excelencia Severo Ochoa No.~CEX2020-001007-S, funded by MCIN/AEI/10.13039/501100011033, by ``ERDF A way of making Europe'', and by FEDER, UE, and by the European Commission -- NextGenerationEU, through Momentum CSIC Programme: Develop Your Digital Talent. This project has also received funding from the European Union's Horizon Europe research and innovation programme under the Marie Sk\l{}odowska-Curie grant agreement No.\ 101106243. The authors gratefully acknowledge the Gauss Centre for Supercomputing e.V. (\url{www.gauss-centre.eu}) for funding this project by providing computing time on the GCS Supercomputer SuperMUC-NG at Leibniz Supercomputing Centre (\url{www.lrz.de}) and the computing time on the high-performance computer "Lise" at the NHR center NHR@ZIB. This center is jointly supported by the Federal Ministry of Education and Research and the state governments participating in the NHR (\url{www.nhr-verein.de}). We furthermore acknowledge the computer resources provided by DESY Zeuthen (PAX cluster) and thank the staff for their support. We are grateful to our colleagues in the CLS initiative for producing the large-volume gauge field configuration ensembles used in this study.

\bibliographystyle{JHEP}
\bibliography{ref.bib}

\end{document}